\documentclass[conference]{IEEEtran}
\usepackage[left=0.70in, right=0.70in, top=1in, bottom=1in]{geometry}
\IEEEoverridecommandlockouts
\usepackage[pdftex]{graphicx}
\usepackage[ruled]{algorithm2e}
\usepackage{algorithmic}
\usepackage[small]{caption}
\usepackage{float}
\usepackage{xspace}
\usepackage{fancyhdr}
\usepackage{amssymb,amsmath}
\usepackage{subfig}
\usepackage{xcolor,amsmath}
\usepackage{epstopdf}
\usepackage{bbding}
\usepackage{cite} 
\usepackage{color}

\begin{document}
\title{Energy-Efficient RIS-Enabled NOMA Communication for 6G LEO Satellite Networks}
\author{Wali Ullah Khan, \textit{Member, IEEE,} Eva Lagunas, \textit{Senior Member, IEEE,} Asad Mahmood, \\ Symeon Chatzinotas, \textit{Fellow, IEEE,}  Bj\"orn Ottersten, \textit{Fellow, IEEE}  \\Interdisciplinary Centre for Security, Reliability and Trust (SnT), University of Luxembourg\\
\{waliullah.khan, eva.lagunas, asad.mahmood, symeon.chatzinotas, bjorn.ottersten\}@uni.lu
}%

\maketitle
\begin{abstract}
Reconfigurable Intelligent surfaces (RIS) have the potential to significantly improve the performance of future 6G LEO satellite networks. In particular, RIS can improve the signal quality of ground terminal, reduce power consumption of satellite and increase spectral efficiency of overall network. This paper proposes an energy-efficient RIS-enabled NOMA communication for LEO satellite networks. The proposed framework simultaneously optimizes the transmit power of ground terminals at LEO satellite and passive beamforming at RIS while ensuring the quality of services. Due to the nature of the considered system and optimization variables, the problem of energy efficiency maximization is formulated as non-convex. In practice, it is very challenging to obtain the optimal solution for such problems. Therefore, we adopt alternating optimization methods to handle the joint optimization in two steps. In step 1, for any given phase shift vector, we calculate efficient power for ground terminals at satellite using Lagrangian dual method. Then, in step 2, given the transmit power, we design passive beamforming for RIS by solving the semi-definite programming. To validate the proposed solution, numerical results are also provided to demonstrate the benefits of the proposed optimization framework.
\end{abstract}



\section{Introduction}
Although 5G networks are still being deployed, researchers are already exploring the potential of 6G technologies, which are expected to offer even faster data speeds, high energy efficiency, lower latency, and more reliable communication \cite{zhang20196g}. These technologies will rely on both terrestrial and non-terrestrial networks, including infrastructure located both on and off the Earth's surface \cite{azari2022evolution}. Low Earth Orbit (LEO) satellite communications, in which satellites orbit at altitudes of 500 to 2000 km, have recently received significant research attention due to their potential to provide worldwide wireless access with low latency \cite{su2019broadband}. LEO satellites can be designed to provide high-bandwidth, low-latency communication, which is particularly important for applications such as real-time video conferencing and online gaming \cite{you2020massive}. 

Despite the advantages mentioned above, there are also some potential challenges to consider. For instance, deploying a large number of low LEO satellites will require a significant amount of spectrum and energy resources \cite{kodheli2020satellite}. To address this, two possible technologies are non-orthogonal multiple access (NOMA) and Reconfigurable Intelligent Surfaces (RIS). NOMA has the capability to solve the problem of spectrum scarcity in satellite communication by accommodating multiple ground users over the same spectrum resource simultaneously \cite{gao2021sum}. On the other hand, RIS has the potential to significantly enhance the performance of LEO satellite networks in terms of coverage and energy efficiency \cite{khan2022opportunities}. RIS consists of a large number of sub-wavelength-sized elements that can be electronically controlled to reflect and manipulate electromagnetic waves. 

In the context of LEO satellite networks, RIS can be used to improve communication between satellites and ground stations. By placing RIS strategically, the incident signals can be efficiently reflected in a specific direction, creating a high-gain antenna. This can improve signal strength and reduce interference, thereby increasing the capacity, coverage and energy consumption of the system. Moreover, since LEO satellite networks require a large number of satellites to provide global coverage, the use of RIS can help reduce the number of satellites needed to achieve the same level of coverage. By improving the efficiency of communication links, RIS can help reduce the cost and complexity of deploying and maintaining LEO satellite networks.

Recently, some researchers have integrated RIS with satellite communications, for example, Dong {\em et al.} \cite{dong2021towards} have maximized the weighted sum rate of integrated terrestrial satellite networks by simultaneously optimising the transmit beamforming at the base station and phase shift matrix at RIS. The authors of \cite{zheng2022ris} have provided a new framework to enhance the average throughput of RIS-enabled LEO networks by jointly optimizing the orientation and passive beamforming at RIS. Lee {\em et al.} \cite{lee2021performance} have jointly optimized the active and passive beamforming to maximize the received signal-to-noise ratio in RIS-enabled LEO satellite networks. Moreover, Khan {\em et al.} \cite{khan2022ris} have proposed a sustainable framework for maximizing the spectral efficiency of RIS-enabled GEO satellite networks by jointly optimizing the transmit power of the satellite and phase shift at RIS. The work in \cite{zheng2022intelligent} have designed a new architecture to enhance the overall channel gain in RIS-enabled LEO satellite networks by jointly optimising the transmit and receive beamforming. Further, Tekbiyik {\em et al.} \cite{tekbiyik2022reconfigurable} have investigated bit error rate and achievable rate for RIS-enabled terahertz communication in LEO satellite networks. Xu {\em et al.} \cite{xu2021intelligent} have also studied a physical layer security problem in RIS-enabled cognitive radio satellite-terrestrial integrated networks. Of late, the author of \cite{dong2021towards} have extended their work in \cite{dong2022intelligent} to maximize the weighted sum rate of RIS-enabled LEO satellite networks by jointly optimizing user scheduling, coordinated transmit beamforming and phase shift design.

It can be observed that the above-existing literature considered RIS-enabled satellite networks adopting orthogonal spectrum techniques. To the best of our knowledge, no work exists on RIS-enabled NOMA LEO satellite communication; hence, it is an open topic to study. Therefore, this paper proposes an optimization framework for RIS-enabled NOMA LEO satellite communication. In particular, we maximize the achievable energy efficiency of the system by simultaneously optimizing the transmit power of LEO satellite and passive beamforming at RIS while ensuring the quality of services. The remainder of the paper is structured as follows. Section II studies the system model of RIS-enabled NOMA LEO satellite communication. Section III provided the proposed energy-efficient solution. Section IV provided numerical results and discussion while Section V concludes this paper.
\begin{figure}[t]
\centering
\includegraphics [width=0.40\textwidth]{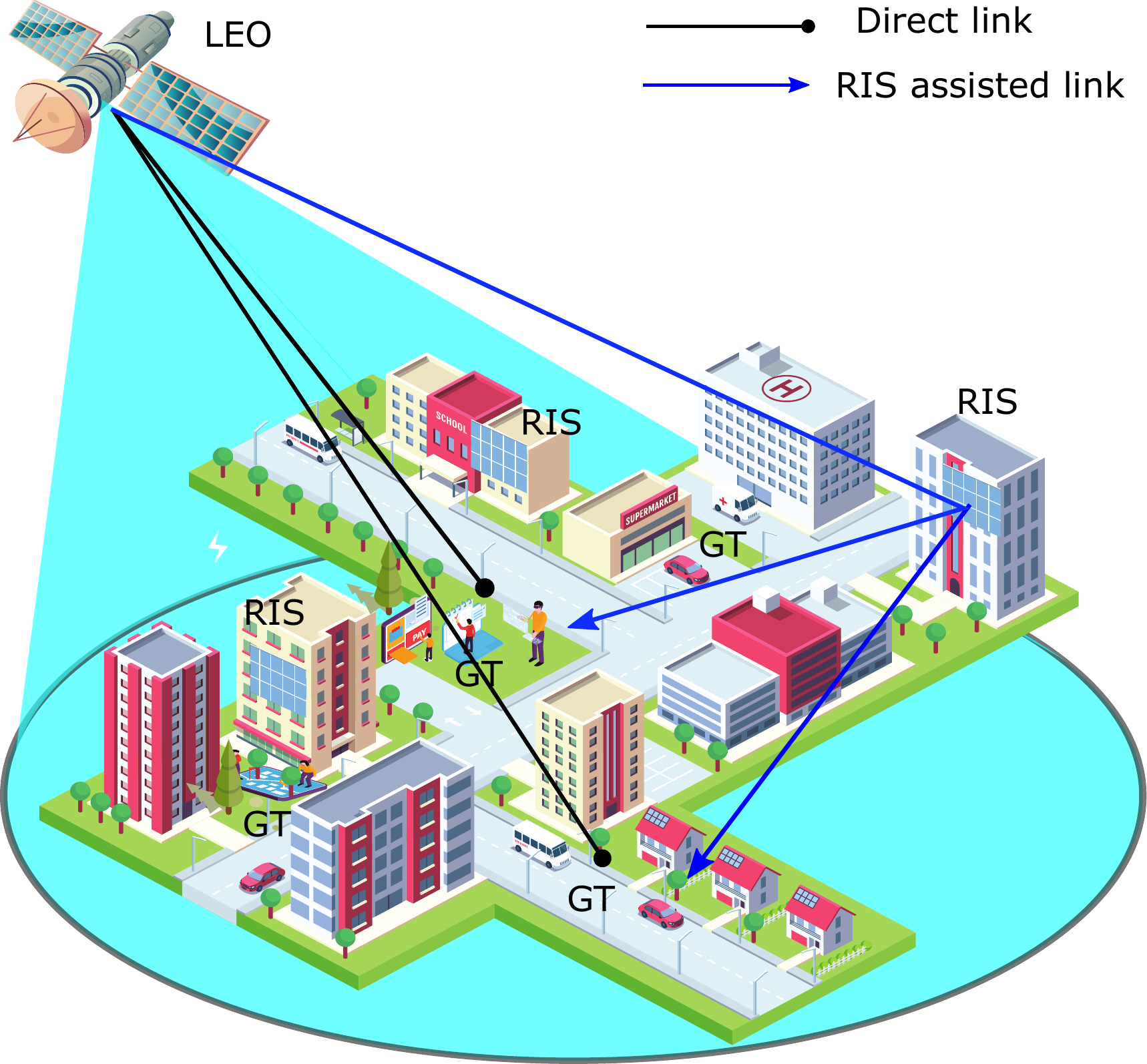}
\caption{System model}
\label{blocky}
\end{figure}
\section{System Model and Problem Formulation}
Consider an LEO satellite that communicates with ground terminals (GTs) in urban areas using the NOMA protocol, as shown in Fig. \ref{blocky}. The LEO satellite uses Ka-band and multi-beam technology. Although NOMA has the potential to accommodate multiple users over the same resource block, we assume that the LEO satellite accommodates only two GTs at any given time for simplicity. However, our work can easily be extended to multiple resource blocks and GTs. We consider that the GTs are mobile and may experience large-scale fading, resulting in performance degradation. Therefore, we also consider using RIS, strategically mounted on building walls, to assist with the signal delivery of NOMA GTs from the LEO satellite. It is assumed that the GTs are equipped with one antenna, RIS consists of multiple reconfigurable elements, and the channel state information is available.

Let us denote the LEO satellite as $l$, the GTs as GT $i$ and GT $j$, respectively. Moreover, the reconfigurable elements are defined as $\mathcal M=\{m|1,2,3,\dots,M\}$, where $m$ indexes a single element. Accordingly, the $M\times M$ diagonal matrix of RIS can be defined as $\boldsymbol\Theta=\text{diag}\{\alpha_1,\alpha_2,\alpha_3,\dots, \alpha_M\}$, where $\alpha_m$ denotes the reconfigurable coefficent of element $m$ and satisfy $|\alpha_m|=1, \ \forall m\in \{1,2,3,\dots, M\}$. Due to different geographical locations, we assume that the channel gains of GTs are not similar. Thus, the direct channel gains from LEO satellite to GTs can be sorted as $h_{l, i}\geq h_{l,j}$ without loss of generality. By considering the block faded channel model, $h_{l,\iota}\in\{i,j\}$ can be expressed as $h_{l,\iota}=\hat{h}_{l,\iota}e^{\hat j\pi\zeta_{l,\iota}}$, where $\hat{h}_{l,\iota}$ denotes the complex-valued channel gain, $\zeta_{l,\iota}$ is the Doppler shift, and $\hat j=\sqrt{-1}$. Considering small and large-scale fading, the complex-valued channel gain can be stated as 
\begin{align}
\small
\bar h_{l,\iota}=\sqrt{G_lG_{l,\iota} \Big(\frac{c}{4\pi f_c d_{l,\iota}}\Big)^2}
\end{align}
where $c$ represents the speed of light, $f_c$ is the carrier frequency, $d_{l,\iota}$ denotes the distance from LEO satellite to GT, $G_{l,\iota}$ shows the received antenna gain, and $G_{l}$ denotes the transmit antenna gain, respectively. Note that the transmit antenna gain mostly depends on radiation pattern and GT geographical location, which can be approximated as
\begin{align}
 G_{l} =   G_{max}\left[\frac{J_1(\vartheta_{l,\iota})}{2\vartheta_{l,\iota}}+36\frac{J_3(\vartheta_{l,\iota})}{\vartheta^3_{l,\iota}}\right]^2,
\end{align}
where $G_{max}$ represents the maximum gain observed at the beam centre, $\vartheta_{l,\iota}=2.07123\sin{(\theta_{l,\iota})}/\sin(\theta_{3dB})$ such that $\theta_{l,\iota}$ denotes the angle between GT and the centre of LEO satellite beam for a given GT location. The angle of 3 dB loss related to the beam's centre is given by $\theta_{3dB}$. Moreover, $J_1$ and $J_2$ are the first-kind Bessel functions having orders 1 and 2. Let us assume that $x_l$ is the superimposed signal of LEO satellite for GT $i$ and GT $j$ such that $x_l=\sqrt{P_l\varrho_{l,i}}x_{l,i}+\sqrt{P_l\varrho_{l,j}}x_{l,j}$, where $P_l$ is the transmit power of LEO satellite while $\varrho_{l,i}$ and $\varrho_{l,j}$ are the power allocation coefficients. Further, $x_{l,i}$ and $x_{l,j}$ are the unit power signals. The signals that GT $i$ and GT $j$ receive from the LEO satellite can be written as
\begin{align}
y_{l,i}&=(h_{l,i}+\boldsymbol{g}_{l,m}\boldsymbol{\Theta}\boldsymbol{f}_{m,i})x_l+\omega_{l,i},\label{y1}
\\
y_{l,j}&=(h_{l,j}+\boldsymbol{g}_{l,m}\boldsymbol{\Theta}\boldsymbol{f}_{m,j})x_l+\omega_{l,j},\label{y2}
\end{align}
where $\omega_{l,\iota}$ is the additive white Gaussian noise (AWGN), $\boldsymbol{g}^H_{l,m}\in M\times1$ denotes the channel gains from LEO satellite to RIS, which can be modelled similar as for the LEO to GTs; however, we omit the detail here for simplicity. Moreover, $\boldsymbol{f}_{m,\iota}=\hat{f}_{m,\iota}d^{-\beta/2}_{l,\iota}\in M\times1$ represent the channel gains from RIS to GT $\iota$, where $\hat{f}_{m,\iota}$ is the Raleigh fading coefficient, $d_{l,\iota}$ is the distance from RIS to GTs, and $\beta$ denotes the pathloss exponent.
Following the downlink NOMA protocol, GT with a higher received channel gain can apply SIC decoding technique to subtract the signal of another user before decoding the desired signal. However, the GT with a lower received channel gain cannot apply SIC and decode the desired signal by treating the signal of other GT as a noise. Based on this observation, the signal-to-noise ratio of GT $i$ can be expressed as 
\begin{align}
\small
\gamma_{l,i}=P_l\varrho_{l,i}|h_{l,i}+\boldsymbol{g}_{l,m}\boldsymbol{\Theta}\boldsymbol{f}_{m,i}|^2/\sigma^2,\label{7}
\end{align}
where $\sigma^2$ is the variance of AWGN. Accordingly, the signal-to-interference plus noise ratio of GT $j$ can be written as
\begin{align}
\gamma_{l,j}=\frac{P_l\varrho_{l,j}|h_{l,j}+\boldsymbol{g}_{l,m}\boldsymbol{\Theta}\boldsymbol{f}_{m,j}|^2}{\sigma^2+P_l\varrho_{l,i}|h_{l,j}+\boldsymbol{g}_{l,m}\boldsymbol{\Theta}\boldsymbol{f}_{m,j}|^2},\label{8}
\end{align}
where the second term in the denominator is the interference of GT $i$. Now we can define the data rate of GT $i$ and GT $j$ from the LEO satellite as
\begin{align}
R_{l,i} = \log_2(1+\gamma_{l,i}),\label{9}
\end{align}
\begin{align}
R_{l,j} = \log_2(1+\gamma_{l,j}),\label{10}
\end{align}

The objective of this work is to maximize the achievable energy efficiency of NOMA LEO satellite system. In this work, the total energy efficiency can be defined as the total achievable capacity divided by the system's total power consumption. This objective can be achieved by optimising the transmit power of GTs at the LEO satellite and passive beamforming of RIS subject to the quality of services constraints. The optimization problem can be formulated as
\begin{alignat}{2}
&\underset{{\boldsymbol{\varrho}, \boldsymbol{\Theta}}}{\text{max}} \ \frac{\log_2(1+\gamma_{l,i})+\log_2(1+\gamma_{l,j})}{P_l(\varrho_{l,i+\varrho_{l,j}})+p_c}\label{OP} \\
 s.t.\ &    \gamma_{l,i}\geq \gamma_{min}, \tag{9a}\label{9a}\\
&  \gamma_{l,j}\geq \gamma_{min},\tag{9b} \label{9b}\\
& 0\leq P_l(\varrho_{l,i}+\varrho_{l,j})\leq P_T,\tag{9c} \label{9c}\\
& |\alpha_m|=1,\ \forall m\in M,\tag{9d} \label{9d}\\
& \varrho_{l,i}+\varrho_{l,j}\leq 1,\tag{9e} \label{9e}\\
 & 0\leq\varrho_{l,i}\leq1, 0\leq\varrho_{l,j}\leq1,\tag{9f} \label{9f}
\end{alignat}
where the objective in (\ref{OP}) is to maximize the achievable energy efficiency of RIS-enabled NOMA LEO satellite network. The constraints (9a) and (9b) guarantee the quality of services for GT $j$ and GT $j$. Constraint (9c) controls the power transmission of the LEO satellite. Constraint (9d) invokes the phase shift of RIS system. Constraints (9e) and (9f) allocate power to GT $i$ and GT $j$ based on downlink NOMA. 

\section{Proposed Energy-Efficient Solution}
The optimization problem in (\ref{OP}) is non-convex due to several factors, i.e., it couples with two variables, interference in rate expression and the fractional objective function. Thus, obtaining a joint optimal solution is challenging because of its high complexity. To efficiently solve this problem, we exploit an alternating optimization approach. Based on this method, the optimization problem in (\ref{OP}) can be efficiently solved in two steps. In the first step, the transmit power of GT $i$ and GT $j$ can be calculated at the LEO satellite for any given passive beamforming at RIS. Then in the second step, the passive beamforming at RIS is computed given the transmit power at the LEO satellite.
\begin{figure*}[!t]
\begin{align}
&\mathcal L(\varrho_{l,i},\varrho_{l,j},\boldsymbol{\lambda}) = (\Psi_{l,i}\log_2(\gamma_{l,i})+\Omega_{l,i}+\Psi_{l,j}\log_2(\gamma_{l,j})+\Omega_{l,j})-\phi^{t-1}(P_l(\varrho_{l,i+\varrho_{l,j}})+p_c)\nonumber\\&+\lambda_{1}\left(P_l\varrho_{l,i}|h_{l,i}+\boldsymbol{g}_{l,m}\boldsymbol{\Theta}\boldsymbol{f}_{m,i}|^2-\gamma_{min}(\sigma^2)\right)+\lambda_{2}(P_l\varrho_{l,j}|h_{l,j}+\boldsymbol{g}_{l,m}\boldsymbol{\Theta}\boldsymbol{f}_{m,j}|^2-\gamma_{min}(\sigma^2\nonumber\\&+P_l\varrho_{l,i}|h_{l,j}+\boldsymbol{g}_{l,m}\boldsymbol{\Theta}\boldsymbol{f}_{m,j}|^2)) +\lambda_3(P_T-P_l\varDelta )+\lambda_4(1-(\varrho_{l,i}+\varrho_{l,j})), \tag{13}\label{250}
\end{align}\hrulefill
\begin{align}
&\dfrac{L(\varrho_{l,i},\varrho_{l,j},\boldsymbol{\lambda})}{\partial \varrho_{l,i}}=\varrho_{l,i}^2 \Big(-O_{l,j} P_l (-\lambda_1 O_{l,i} P_l+(\lambda_4+P_l(\lambda_3+\phi^{t-1}+\lambda_2 O_{l,j}\gamma_{min}))\sigma^2)\Big)+\varrho_{l,i}\Big(-\sigma(-\lambda_1 O_{l,i} P_l\nonumber\\& +(\lambda_4 +(\lambda_3+\phi^{t-1})P_l)\sigma^2+O_{l,j} P_l (-\Psi_{l,i}+\Psi_{l,j}+\lambda_2\gamma_{min}\sigma^2))\Big) +\Psi_{l,i}\sigma^2=0. \tag{15}\label{251}
\end{align}\hrulefill
\begin{align}
A&=\sigma^2 ((-\lambda_1 O_{l,i} P_l+(\lambda_4+(\lambda_3+\phi^{t-1})\sigma^2)+O_{l,j} P_l(-\Psi_{l,i}+\Psi_{l,j}+\lambda_2 \gamma_{min}\sigma^2)),\nonumber\\
B&=\sigma^4(4 O_{l,j} P_l \Psi_{l,i}(-\lambda_1 O_{l,i} P_l+(\lambda_4+P_l(\lambda_3+\phi^{t-1}+\lambda_2 O_{l,j} \gamma_{min}))\sigma^2)\nonumber\\&+((-\lambda_1 O_{l,i} P_l+(\lambda_4+(\lambda_3+\phi^{t-1})P_l)\sigma^2)+O_{l,j} P_l(-\Psi_{l,i}+\Psi_{l,j}+\lambda_2\gamma_{min}\sigma^2))^2),\tag{17}\label{252}\\
C&=-2 O_{l,j} P_l (-\lambda_1 O_{l,i} P_l+(\lambda_4+P_l (\lambda_3+\phi^{t-1}+\lambda_2O_{l,j} \gamma_{min}))\sigma^2).\nonumber 
\end{align}\hrulefill
\end{figure*}
\subsection{NOMA Power Allocation at LEO Satellite: Step-1}
For any given passive beamforming $\boldsymbol{\Theta}$ at RIS, the problem in (\ref{OP}) can be simplified as power allocation optimization at LEO satellite, such as
\begin{alignat}{2}
&\underset{\varrho_{l,i},\varrho_{l,j}}{\text{max}}\ \frac{\log_2(1+\gamma_{l,i})+\log_2(1+\gamma_{l,j})}{P_l(\varrho_{l,i+\varrho_{l,j}})+p_c}\label{OP1} \\
 s.t.\ & P_l\varrho_{l,i}|h_{l,i}+\boldsymbol{g}_{l,m}\boldsymbol{\Theta}\boldsymbol{f}_{m,i}|^2\geq \gamma_{min} (\sigma^2), \tag{10a}\label{10a}\\
 & P_l\varrho_{l,j}|h_{l,j}+\boldsymbol{g}_{l,m}\boldsymbol{\Theta}\boldsymbol{f}_{m,j}|^2\geq \gamma_{min} (\sigma^2+P_l\varrho_{l,i}\nonumber\\&|h_{l,j}+\boldsymbol{g}_{l,m}\boldsymbol{\Theta}\boldsymbol{f}_{m,j}|^2),\ (\ref{9c}), (\ref{9e}).\tag{10b}\label{10b}
\end{alignat}
The proposed optimization in (\ref{OP1}) is a fractional problem due to the objective function. It can be efficiently transformed into a non-fractional, which can be expressed as
\begin{alignat}{2}
&\underset{\varrho_{l,i},\varrho_{l,j}}{\text{max}}\ \log_2(1+\gamma_{l,i})+\log_2(1+\gamma_{l,j})\nonumber\\&-\phi^{t-1}(P_l(\varrho_{l,i+\varrho_{l,j}})+p_c)\label{OP12} \\
 s.t.\ & (\ref{10a}),(\ref{10b}), (\ref{9c}),(\ref{9e}), \nonumber
\end{alignat}
where $\phi$ denotes a non-negative parameter and $t$ indexes iteration number. To solve problem (\ref{OP12}), the $\boldsymbol{\gamma}$ and $\boldsymbol{\varrho}$ can be updated in each iteration as $\phi^t=(\log_2(1+\gamma_{l,i})+\log_2(1+\gamma_{l,j}))/(P_l(\varrho_{l,i}+\varrho_{l,j})+p_c)$. Moreover, the maximum energy efficiency in (\ref{OP12}) can be computed as $\eta=(\log_2(1+\gamma^t_{l,i})+\log_2(1+\gamma^t_{l,j}))-\phi^{t-1}(P_l(\varrho^t_{l,i}+\varrho^t_{l,j})+p_c)$. During the iterative process, the value of $\phi$ increases while the value of $\eta$ reduces and approaches to zero. The maximum energy efficiency is achieved when $\phi$ is maximum and $\eta=0$.

Next, it is important to see that how the optimization in (\ref{OP12}) can be solved for a given $\phi$. It is clear that the problem (\ref{OP12}) is non-convex due to the non-concave objective function. 
To simplify the problem and make it more manageable, we use the successive convex approximation method.
By applying this method, the sum capacity in the objective of problem (\ref{OP12}) can be expressed as
$\Psi_{l,i}\log_2(\gamma_{l,i})+\Omega_{l,i}+\Psi_{l,j}\log_2(\gamma_{l,j})+\Omega_{l,j}$ with
the values of
$\Psi_{l,\iota}=\frac{\gamma_{l,\iota}}{1+\gamma_{l,\iota}}$ and $\Omega_{l,\iota}=\log_2(1+\gamma_{l,\iota})-\Psi_{l,\iota}\log_2(\gamma_{l,\iota}),\ \iota\in\{i,j\}$.
Now the optimization problem in (\ref{OP12}) can be reformulated as
\begin{alignat}{2}
&\underset{\varrho_{l,i},\varrho_{l,j}}{\text{max}}(\Psi_{l,i}\log_2(\gamma_{l,i})+\Omega_{l,i}+\Psi_{l,j}\log_2(\gamma_{l,j})+\Omega_{l,j})\nonumber\\&-\phi^{t-1}(P_l(\varrho_{l,i+\varrho_{l,j}})+p_c)\label{OP13} \\
 s.t.\ & (\ref{10a}),(\ref{10b}), (\ref{9c}),(\ref{9e}), \nonumber
\end{alignat}
Now we employ Lagrangian dual method to efficiently solve the optimization problem (\ref{OP13}). The Lagrangian function to solve problem (\ref{OP13}) can be defined as Equation (\ref{250}),
where $\boldsymbol{\lambda}=\{\lambda_1,\lambda_2,\lambda_3,\lambda_4\}$ and $\varDelta=\varrho_{l,i}+\varrho_{l,j}$. Now we use Karush-KhunTucker (KKT) conditions to (\ref{250}) such as
\begin{align}
\frac{\partial \mathcal L(\varrho_{l,i},\varrho_{l,j},\boldsymbol{\lambda})}{\partial \varrho_{l,i},\varrho_{l,j}}|_{\varrho_{l,i},\varrho_{l,j}=\varrho_{l,i}^*,\varrho_{l,j}^*}=0,\tag{14}
\end{align}
After the straightforward derivative of (\ref{250}), we obtain a polynomial, which can be written as Equation (\ref{251}).
Now solving $\varrho_{l,i}$, it can be expressed as
\begin{align}
 \varrho^*_{l,i}=\dfrac{A \pm \sqrt{B}}{C}  \tag{16} 
\end{align}
where the values of $A$, $B$, and $C$ can be found in Equation (\ref{252}).
In addition, the values of $O_{l,i}$ and $O_{l,j}$ can be given as
\begin{align}
&O_{l,i}=|h_{l,i}+\boldsymbol{g}_{l,m}\boldsymbol{\Theta}\boldsymbol{f}_{m,i}|^2.\tag{18}\\
&O_{l,j}=|h_{l,j}+\boldsymbol{g}_{l,m}\boldsymbol{\Theta}\boldsymbol{f}_{m,j}|^2.\tag{19}
\end{align}
After computing $\varrho^*_{l,i}$, the value of $\varrho^*_{l,j}$ can be efficiently achieve as
\begin{align}
\varrho^*_{l,j}=1-\varrho^*_{l,i}. \tag{20}
\end{align}

\begin{figure*}[!t]
\begin{align}
&\log_2(1+\text{Tr}(\boldsymbol{\Xi}\boldsymbol{G}_{l,i})/\sigma^2)+\log_2\bigg(1+\frac{\text{Tr}(\boldsymbol{\Xi}\boldsymbol{G}_{l,j})}{\text{Tr}(\boldsymbol{\Xi}\boldsymbol{\bar{G}}_{l,j})+\sigma^2}\bigg)=(\log_2(\sigma^2+\text{Tr}(\boldsymbol{\Xi}\boldsymbol{G}_{l,i}))-\log_2(\sigma^2))\nonumber\\
&+(\log_2(\text{Tr}(\boldsymbol{\Xi}\boldsymbol{\bar{G}}_{l,j})+\sigma^2+\text{Tr}(\boldsymbol{\Xi}\boldsymbol{G}_{l,j}))-\log_2(\text{Tr}(\boldsymbol{\Xi}\boldsymbol{\bar{G}}_{l,j})+\sigma^2)). \tag{28}\label{253}
\end{align}\hrulefill
\end{figure*}
\subsection{Passive Beamforming at RIS: Step-2}
Under given $\varrho^*_{l,i},\varrho^*_{l,j}$ at LEO satellite, the passive beamforming at RIS can be further simplified. We can observe that the direct links from satellite to GTs have no impact on the passive beamforming. Thus it can be efficiently ignored. The Equations (\ref{7}) and (\ref{8}) can be then reformulated as
\begin{align}
\bar\gamma_{l,i}=P_l\varrho_{l,i}|\boldsymbol{g}_{l,m}\boldsymbol{\Theta}\boldsymbol{f}_{m,i}|^2/\sigma^2,\tag{21}\label{25}\\
\bar\gamma_{l,j}=\frac{P_l\varrho_{l,j}|\boldsymbol{g}_{l,m}\boldsymbol{\Theta}\boldsymbol{f}_{m,j}|^2}{\sigma^2+P_l\varrho_{l,i}|\boldsymbol{g}_{l,m}\boldsymbol{\Theta}\boldsymbol{f}_{m,j}|^2},\tag{22}\label{26}
\end{align}
Then, the optimization for passive beamforming at RIS can be formulated as
\begin{alignat}{2}
&\underset{\boldsymbol{\Theta}}{\text{max}}\ \log_2(1+\bar\gamma_{l,i})+\log_2(1+\bar\gamma_{l,j})\tag{23}\label{OP2} \\
 s.t.\  &    \bar\gamma_{l,i}\geq \bar\gamma_{min}, \tag{23a}\label{23a}\\
 & \bar\gamma_{l,j}\geq \bar\gamma_{min},\tag{23b}\label{23b} \\
   & |\alpha_m|=1, m\in M, \tag{23c}\label{23c}
\end{alignat}
 To easily handle the terms $|\boldsymbol{g}_{l,m}\boldsymbol{\Theta}\boldsymbol{f}_{m,i}|^2$ and $|\boldsymbol{g}_{l,m}\boldsymbol{\Theta}\boldsymbol{f}_{m,j}|^2$, let us denote $\boldsymbol{\xi}=[\xi_1,\xi_2,\xi_3,\dots,\xi_{m}]$ is the diagonal vector of reconfigurable elements in passive beamforming matrix $\boldsymbol{\Theta}$, where $\xi_m=\alpha^H_m$. Then, we introduce an auxiliary vector $\boldsymbol{\hat{H}}_{l,\iota}$ such that $\boldsymbol{\hat{H}}_{l,\iota}=\boldsymbol{g}_{l,m}\circ\boldsymbol{f}_{m,\iota}$, where $\circ$ represents the Hadamard product and $\iota\in\{i,j\}$. Now the following equality can efficiently hold 
as $|\boldsymbol{g}_{l,m}\boldsymbol{\Theta}\boldsymbol{f}_{m,\iota}|^2=|\boldsymbol{\hat{H}}_{l,\iota}|^2$. Adopting these updates, the problem in (\ref{OP2}) can be rewritten as
\begin{alignat}{2}
&\underset{\boldsymbol{\xi}}{\text{max}}  \ \log_2(1+P_l\varrho_{l,i}|\boldsymbol{\xi}^H\boldsymbol{\hat{H}}_{l,i}|^2/\sigma^2)\nonumber\\&+\log_2\bigg(1+\frac{P_l\varrho_{l,j}|\boldsymbol{\xi}^H\boldsymbol{\hat{H}}_{l,j}|^2}{P_l\varrho_{l,i}|\boldsymbol{\xi}^H\boldsymbol{\hat{H}}_{l,j}|^2+\sigma^2}\bigg)\tag{24}\label{OP21} \\
 s.t.\  &   P_l\varrho_{l,i}|\boldsymbol{\xi}^H\boldsymbol{\hat{H}}_{l,i}|^2\geq \bar\gamma_{min}(\sigma^2), \tag{24a}\label{24a}\\
 & P_l\varrho_{l,j}|\boldsymbol{\xi}^H\boldsymbol{\hat{H}}_{l,j}|^2\geq \bar\gamma_{min}(P_l\varrho_{l,i}|\boldsymbol{\xi}^H\boldsymbol{\hat{H}}_{l,j}|^2+\sigma^2), \tag{24b}\label{24b} \\
   & |\xi_m|=1, m\in M,\tag{24c}\label{24c}
\end{alignat}
To solve (\ref{OP21}), we re-expressed $P_l\varrho_{l,\iota}|\boldsymbol{\xi}^H\boldsymbol{\hat{H}}_{l,\iota}|^2$ as
\begin{align}
& P_l\varrho_{l,\iota}|\boldsymbol{\xi}^H\boldsymbol{\hat{H}}_{l,\iota}|^2= \boldsymbol{\xi}^HP_l\varrho_{l,\iota}\boldsymbol{\hat{H}}_{l,\iota}\boldsymbol{\hat{H}}^H_{l,\iota}\boldsymbol{\xi}\nonumber\\& = \text{Tr}(\boldsymbol{\xi}^HP_l\varrho_{l,\iota}\boldsymbol{\hat{H}}_{l,\iota}\boldsymbol{\hat{H}}^H_{l,\iota}\boldsymbol{\xi})=\text{Tr}(P_l\varrho_{l,\iota}\boldsymbol{\hat{H}}_{l,\iota}\boldsymbol{\hat{H}}^H_{l,\iota}\boldsymbol{\xi}\boldsymbol{\xi}^H).\tag{25}
\end{align}
Next, we define two auxiliary matrices as $\boldsymbol{G}_{l,\iota}=P_l\varrho_{l,\iota}\boldsymbol{\hat{H}}_{l,\iota}\boldsymbol{\hat{H}}^H_{l,\iota}$ and $\boldsymbol{\Xi}=\boldsymbol{\xi}\boldsymbol{\xi}^H$. We can observe that $\boldsymbol{G}_{l,\iota}$ and $\boldsymbol{\Xi}$ are semi-definite. Now the problem in (\ref{OP21}) can be reformulated as
\begin{alignat}{2}
&\underset{\boldsymbol{\Xi}}{\text{max}} \ \log_2(1+\text{Tr}(\boldsymbol{\Xi}\boldsymbol{G}_{l,i})/\sigma^2)\nonumber\\&+\log_2\bigg(1+\frac{\text{Tr}(\boldsymbol{\Xi}\boldsymbol{G}_{l,j})}{\text{Tr}(\boldsymbol{\Xi}\boldsymbol{\bar{G}}_{l,j})+\sigma^2}\bigg)\tag{26}\label{OP22} \\
 s.t.\  &  \text{Tr}(\boldsymbol{\Xi}\boldsymbol{G}_{l,i})\geq \bar\gamma_{min}(\sigma^2), \tag{26a}\label{26a}\\
 & \text{Tr}(\boldsymbol{\Xi}\boldsymbol{G}_{l,j})\geq \bar\gamma_{min}(\text{Tr}(\boldsymbol{\Xi}\boldsymbol{\bar{G}}_{l,j})+\sigma^2), \tag{26b}\label{26b} \\
   & \text{diag}\{\boldsymbol{\Xi}\}=1, \tag{26c} \label{26c}\\
& \boldsymbol{\Xi}\succeq 1, \tag{26d}\label{26d}\\
& \text{rank}(\boldsymbol{\Xi}) = 1, \tag{26e}\label{26e}
\end{alignat}
where $\boldsymbol{\bar{G}}_{l,j}=P_l\varrho_{l,i}\boldsymbol{\hat{H}}_{l,j}\boldsymbol{\hat{H}}^H_{l,j}$. The $\text{diag}\{\boldsymbol{\Xi}\}$ is used to return the diagonal values of $\boldsymbol{\Xi}$ which means $\text{diag}\{\boldsymbol{\Xi}\}=1$ equals to $\xi^2_m=1$ and $|\xi_m|=1$.

The optimization in (\ref{OP22}) is still non-convex due to the objective function and the rank one constraint. Let us start with the rank one constraint in (\ref{26e}), which can be efficiently replaced by a semi-definite (convex) constraint as $\boldsymbol{\Xi}-\boldsymbol{\hat{\xi}}\boldsymbol{\hat{\xi}}^H\succeq 0$, where $\boldsymbol{\hat{\xi}}$ denotes a vector of auxiliary variables. Now we can replaced $\boldsymbol{\Xi}-\boldsymbol{\hat{\xi}}\boldsymbol{\hat{\xi}}^H\succeq 0$ by convex Schur complement as
\begin{align}
\begin{bmatrix}
\boldsymbol{\Xi} & \boldsymbol{\hat{\xi}} \\
\boldsymbol{\hat{\xi}}^H & 1
\end{bmatrix} \succeq 0.\tag{27}
\end{align}
Next, we take the objective function of the problem (\ref{OP22}), which can be efficiently written as Equation (\ref{253}).
It can be observed that (\ref{253}) is a DC function and be handled by DC programming. However, the $\boldsymbol{\Xi}$ is a complex matrix and calculating its partial derivation through the traditional way is impossible. Thus, it is necessary to calculate its derivation for both imaginary and real parts. After incorporating the above changes, the problem (\ref{OP22}) becomes semi-definite programming, which is convex. This problem can be efficiently handled through any standard convex solver such as CVX.

 
         
    

\begin{figure}[!t]
\centering
\includegraphics [width=0.40\textwidth]{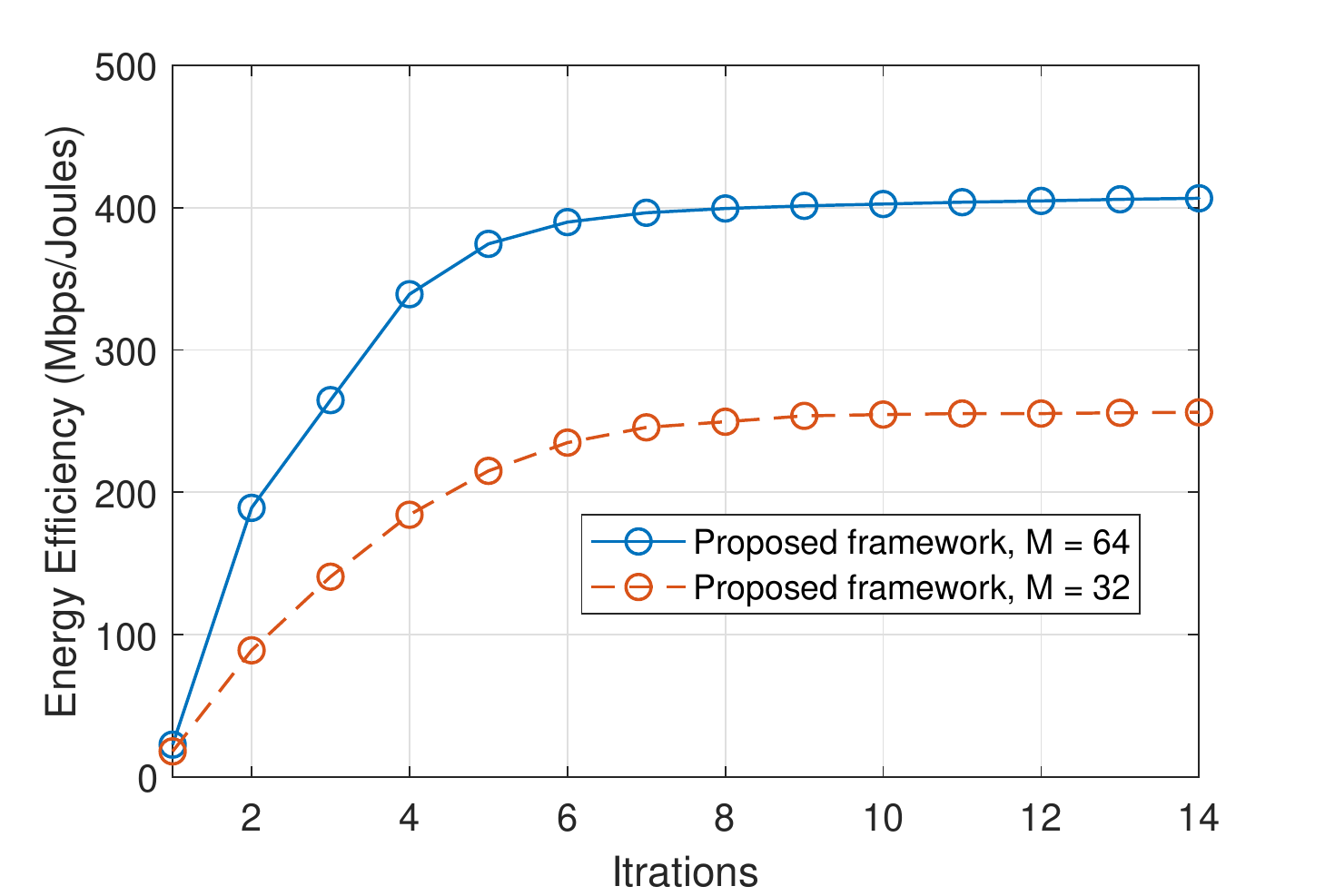}
\caption{Convergence of the proposed optimization framework.}
\label{Fig1VTC}
\end{figure}
\section{Results and Discussion}
In this section, we present the numerical results obtained through Monte Carlo simulations. We compare three frameworks: the proposed framework (Section III), the benchmark framework (where the phase shift vector of RIS is fixed and only the transmit power of the LEO satellite is optimized), and the conventional framework (where RIS is not involved in the proposed model). Unless mentioned otherwise, the simulation parameters are set as follows. We consider that the proposed RIS-enabled NOMA LEO satellite network is operating over Ka-band (17.7-19.7 GHz). The number of NOMA GTs is 2, the number of reconfigurable elements at RIS is 64, the maximum transmission power of the LEO satellite is 50 dBm, the quality of services per GT is set as 10 Mbps/Joules, and the bandwidth of each GT is 20 MHz.

Fig. \ref{Fig1VTC} illustrates the convergence behavior of the proposed optimization framework. The plot shows the achievable energy efficiency as a function of the number of iterations for the systems with 34 and 64 reconfigurable RIS elements. The results demonstrate that the proposed framework achieves convergence after only 8 iterations, indicating its low-complexity nature. Furthermore, it can be observed that the system with more reconfigurable RIS elements achieves higher energy efficiency compared to the system with fewer reconfigurable elements. Additionally, the convergence behavior of the proposed framework for both systems is similar, highlighting its suitability for large-scale systems.

\begin{figure}[!t]
\centering
\includegraphics [width=0.40\textwidth]{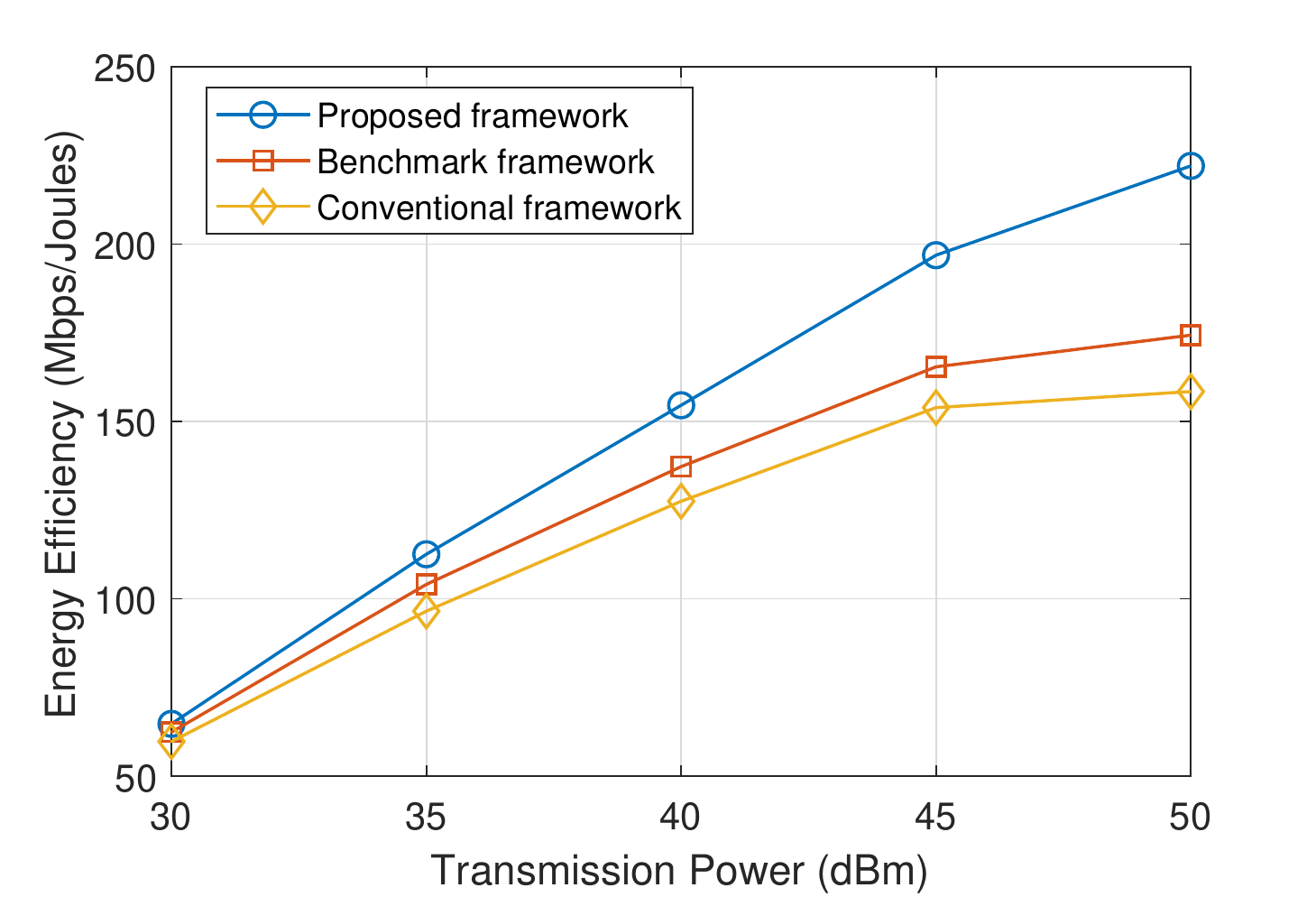}
\caption{Available transmit power versus energy efficiency of the LEO satellite network for $M=32$ and $\gamma_{min}=10$ Mbps/Joules.}
\label{Fig2VTC}
\end{figure}

Figure \ref{Fig2VTC} illustrates the impact of LEO power on the achievable energy efficiency of the satellite network for all three frameworks. The results demonstrate that the achievable energy efficiency of the system increases for all three frameworks as the available transmit power increases. Additionally, we can see that the results of all three frameworks follow a bell-shaped curve, where the achievable energy efficiency exponentially increases with an increase in the transmission power of the LEO satellite and then slows down towards saturation. However, the proposed framework consistently outperforms the other two frameworks for all values of transmit power. Moreover, the gap between the proposed framework and the other frameworks increases as the transmit power increases, indicating the poor performance of the benchmark and conventional frameworks.

\begin{figure}[!t]
\centering
\includegraphics [width=0.40\textwidth]{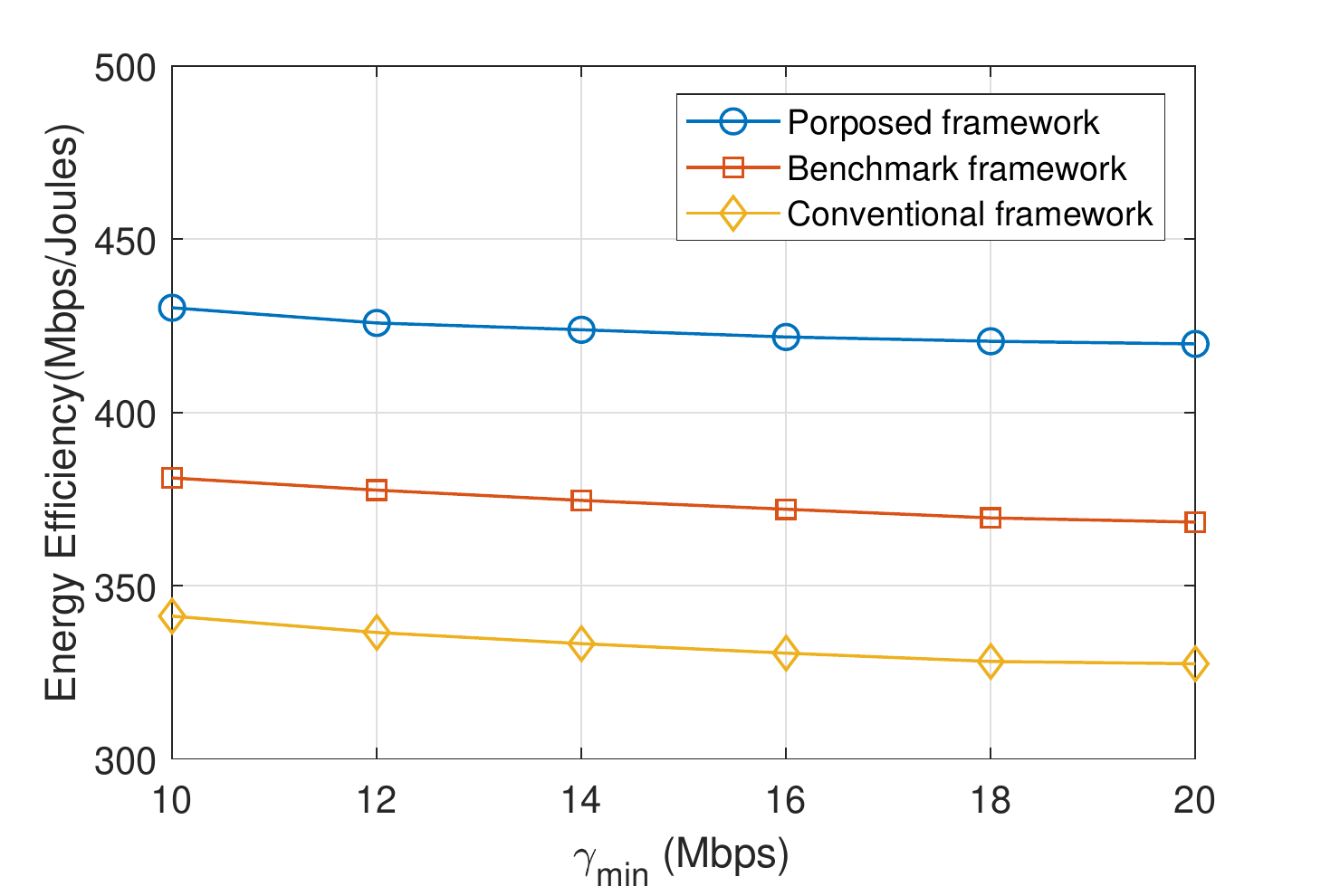}
\caption{Quality of services requirements versus achievable energy efficiency of the LEO satellite network for $P_l=50$ and $M=64$.}
\label{Fig3VTC}
\end{figure}

Finally, Fig. \ref{Fig3VTC} shows the achievable energy efficiency of the system against the increasing values of quality of services requirements. The results show that the achievable energy efficiency of the system slightly decreases for all frameworks as the transmit power of the LEO satellite increases. This is due to the fact that, as the value of $\gamma_{min}$ increases, more power is required for GTs to meet the minimum quality of service requirements, thereby affecting the achievable energy efficiency of the LEO satellite network. However, the proposed framework achieves higher energy efficiency compared to the other two frameworks. Additionally, it can be observed that the benchmark framework performs better than the conventional framework, highlighting the potential benefits of using RIS in satellite networks. 
\section{Conclusions}
In this paper, we present a new framework for optimizing the energy efficiency of RIS-enabled NOMA LEO satellite communication networks. The transmit power of GTs at the LEO satellite and passive beamforming at RIS are simultaneously optimized to maximize the achievable energy efficiency of the satellite network while ensuring the quality of services. The non-convex optimization problem is addressed through alternating optimization in two steps. Firstly, the NOMA power allocation at the LEO satellite is computed by successive convex optimization and dual Lagrangian methods, given the fixed phase shift at RIS. Then, passive beamforming at RIS, a standard convex method, is adopted to handle the semi-definite programming, given the power allocation at the LEO satellite. Numerical results demonstrate that the proposed framework achieves high energy efficiency and converges within a few iterations.

\bibliographystyle{IEEEtran}
\bibliography{Wali_EE}

\end{document}